\documentclass{sig-alternate}
\usepackage{hyperref}
\usepackage{url}
\usepackage{multirow}
\usepackage{tikz}
\usepackage{pgfplots}
\usepackage{graphics}
\usetikzlibrary{decorations.pathmorphing}

\begin{document}

\title{Preliminary Experiments using Subjective Logic for the Polyrepresentation of Information Needs}
%
%
%
%
%

\numberofauthors{3} 
%
\author{
%
%
\alignauthor
Christina Lioma\\
       \affaddr{Computer Science}\\
       \affaddr{University of Copenhagen Denmark}\\
       \email{c.lioma@diku.dk}
\alignauthor
Birger Larsen\\
       \affaddr{Royal School of Library and Information Science Copenhagen Denmark}\\
       \email{blar@iva.dk}
\alignauthor
Peter Ingwersen\titlenote{Second affiliation: Oslo University College, Norway}\\
       \affaddr{Royal School of Library and Information Science Copenhagen Denmark}\\
       \email{pi@iva.dk}
}
\date{30 July 1999}

\maketitle
\begin{abstract}
According to the principle of \textit{polyrepresentation}, retrieval accuracy may improve through the combination of multiple and diverse information object representations about e.g. the context of the user, the information sought, or the retrieval system \cite{Ingwersen:1996,Ingwersen:2005}. Recently, the principle of polyrepresentation was mathematically expressed using \textit{subjective logic} \cite{Josang:2001}, where the potential suitability of each representation for improving retrieval performance was formalised through degrees of belief and uncertainty \cite{Lioma:2010}. No experimental evidence or practical application has so far validated this model.

We extend the work of Lioma et al. (2010) \cite{Lioma:2010}, by providing a practical application and analysis of the model. We show how to map the abstract notions of belief and uncertainty to real-life evidence drawn from a retrieval dataset. We also show how to estimate two different types of polyrepresentation assuming either (a) independence or (b) dependence between the information objects that are combined. We focus on the polyrepresentation of different types of context relating to user information needs (i.e. work task, user background knowledge, ideal answer) and show that the subjective logic model can predict their optimal combination prior and independently to the retrieval process.   
 
\end{abstract}




\section{Introduction}
\label{s:Introduction}

The principle of \textit{polyrepresentation} \cite{Ingwersen:1996,Ingwersen:2005} posits that information retrieval (IR) effectiveness may improve through the consideration of multiple and diverse representations of information objects or processes, such as IR systems, models, or context about the documents and queries and their mediation to the user. This consideration is usually implemented experimentally as combination of diverse representations with respect to some criterion, more commonly retrieval effectiveness. 
Whereas in principle, polyrepresentation aims to use different information object representations that may enhance IR performance, in practice, each representation is often associated with different degrees of uncertainty regarding the enhancement that it may bring. Therefore, if one representation is weaker (less reliable or accurate) than the others, this should be reflected in the combination process, otherwise effectiveness will suffer. In addition, given the potentially high-dimensional and noisy data contained in information object representations, many different combinations could be produced, fetching various different retrieval results. The process of deciding which combination of all is optimal for retrieval is an open problem, which was recently formalised by Lioma et al. (2010) \cite{Lioma:2010} in the context of polyrepresentation using a type of probabilistic logic called subjective logic \cite{Josang:2001}. This formalism was not accompanied by experimental evidence or practical applications.

This work can be seen as an extension of the work of Lioma et al. (2010); we provide a practical application of that model and show that it is experimentally validated. We show how to map the abstract notions of belief and uncertainty to real-life evidence drawn from a retrieval test collection, and how to estimate two different types of combinations for polyrepresentation assuming either (a) independence or (b) dependence between the information objects that are combined. We focus on the polyrepresentation of different types of context relating to user information needs (i.e. work task, user background knowledge, ideal answer) and show that the subjective logic model can predict their optimal combination prior and independently to the retrieval process. This finding holds for six different standard evaluation measures and two state of the art retrieval models. Furthermore, we look at non-commutativity in polyrepresentation in order to trace the bias between dependent contextual representations, and experimentally show its impact to retrieval effectiveness.

The remainder of this work is organised as follows. 
Section \ref{s:RelatedWork} overviews related work. 
Section \ref{s:Model} summarises the model of Lioma et al. (2010). 
Section \ref{s:Experiments} presents the experimental evaluation of the model. 
Section \ref{s:Discussion} discusses our findings and their limitations. 
Section \ref{s:Conclusions} concludes this work.

\section{Related Work}
\label{s:RelatedWork}

There exist several applications of the principle of polyrepresentation to IR (see \cite{Frommholz:2010,Lioma:2010} for overviews). Regarding the application of polyrepresentation to user information needs, that is the focus of this work, most studies use polyrepresentation for query expansion. Diriye et al. (2009) \cite{DiriyeB:2009} apply polyrepresentation for interactive query expansion, i.e. to improve the suggestion of expansion terms to the users during their search in order to enable better retrieval performance. They show how providing supplementary information on expansion terms can address ambiguity and uncertainty issues, and can improve the perceived usefulness of the terms. In a different (non-interactive approach), Efron and Winget (2010) \cite{EfronW10} use polyrepresentation to combine so-called query aspects, which they collect by skimming the top $k$ documents retrieved for a given query. They consider these skimmed documents relevant and build pseudorelevance judgments without human intervention. 

Our study differs from the above: we do not apply polyrepresentation for query expansion, but rather as a means to combine contextual evidence about the queries. This is more reminiscent of evidence combination or opinion fusion approaches, as explained below. The effect of capturing multiple representations of query context from a single user or from multiple users is a topic that has long attracted interest. According to Croft (2000) \cite{Croft00}, McGill et al. (1979) carried out a study of factors affecting retrieval by different users when they were assigned the same information need as a starting point. They found that there was surprisingly little overlap between the documents retrieved by the different users. Saracevic and Kantor (1988) \cite{ASI:ASI4} also found that when different users constructed Boolean queries based on the same descriptions of the information need, there was little overlap in the retrieved sets. One of the earliest perhaps retrieval models that explicitly incorporated the notion of multiple query representations was proposed by Turtle and Croft (1991) \cite{TurtleC91}. Soon after this, Belkin et al. (1993) \cite{BelkinCCC93} carried out a systematic study on the effects of query combination and verified that retrieval effectiveness could be substantially improved by query combination, but that the effectiveness of the combination depends on the effectiveness of the individual queries. Queries that provided less evidence about relevance should optimally have lower weights in the combination, because bad query representations could reduce effectiveness when combined with better representations. 

This is exactly the topic we address in this work, based on the model proposed in Lioma et al. (2010). How to identify and separate `good' from `bad' query representations, so that undesirable combinations can be avoided, considering not only the features of the query representations, but also the way in which they are combined.

\section{Polyrepresentation Using \\ Subjective Logic}
\label{s:Model}
This section outlines basic notions in subjective logic and how it has been applied to polyrepresentation in Lioma et al. (2010). A detailed introduction into subjective logic can be found in J\o sang (2001) \cite{Josang:2001}. 

\subsection{Subjective Logic Preliminaries}
\label{ss:SL-preliminaries}
The starting point is a frame of discernment defined over a proposition. Arguments in subjective logic are \textit{opinions} representing the belief that the proposition is true\footnote{The opinion space is a subset of the belief space used in the Dempster-Shafer belief theory \cite{Smets:1994}.}. An opinion is formally defined as $\omega^A_X$, where $A$ is the opinion owner, and $x$ is the proposition to which the opinion $\omega$ applies. An opinion can be decomposed into $\omega^A_x = (b,d,u,\alpha)$, where $b$ is $A$'s belief that the proposition is true, $d$ is $A$'s disbelief that the proposition is true\footnote{This corresponds to \textit{doubt} in Shafer (1976) \cite{Shafer:1976}.}, $u$ is $A$'s uncertainty about the proposition, and $\alpha$ is an a priori probability in the absence of committed belief mass. 
Given a proposition for which we have an opinion, we can estimate the probability expectation that the proposition is true as:
\begin{equation}
\label{eq:expectation}
E=b+\alpha \cdot u
\end{equation}

If there is one opinion only about the proposition, Equation \ref{eq:expectation} is computed directly from the components of that opinion. If however several opinions exist about a proposition, assessing the truth of the proposition consists in fusing these opinions and estimating Equation \ref{eq:expectation} from their combined components. Subjective logic describes several fusion operators, suitable for different situations. Two of these are outlined next: \textit{consensus}, which combines independent opinions without bias, and \textit{recommendation}, which combines dependent opinions by modelling the influence of one opinion upon the other. 


\paragraph{Consensus} Let us assume two independent opinions about a proposition $x$: $\omega^A_x = (b^A_x,d^A_x,u^A_x,\alpha^A_x)$ and $\omega^B_x = (b^B_x,d^B_x,u^B_x,\alpha^B_x)$. Their consensus  is $\omega^{A \oplus B}_x$ with components: 

\begin{equation}
 \label{eq:consensus1}
b^{{A \oplus B}}_x = \frac{b^A_x u^B_x + b^B_x u^A_x}{\kappa}, \qquad d^{{A \oplus B}}_x = \frac{d^A_x u^B_x + d^B_x u^A_x}{\kappa}
\end{equation}

\begin{equation}
 \label{eq:consensus2}
u^{{A \oplus B}}_x = \frac{u^A_x u^B_x}{\kappa}, \qquad \kappa = u^A_x + u^B_x - u^A_x u^B_x
\end{equation}
\noindent 
This operation\footnote{The consensus operator is similar to Dempster's rule \cite{Dempster:1968} (see \cite{Josang:2001}, Section 5.3, for a discussion on the difference between the two).} is commutative, associative and assumes that not all of the combined opinions have zero uncertainty. %


\paragraph{Recommendation} Let us assume two opinions that are not independent of each other, but where one influences the other. Let $\omega^B_x = (b^B_x, d^B_x, u^B_x, \alpha^B_x)$ be $B$'s opinion about the proposition, and $\omega^A_B = (b^A_B, d^A_B, u^A_B, \alpha^A_B)$ be $A$'s opinion about $B$'s recommendation ($B$'s influence to $A$). Their combination by recommendation $\omega^{A \otimes B}_x$ is $A$'s opinion about the proposition as a result of the recommendation from $B$, with components:
\begin{equation}
 \label{eq:recommendation1}
b^{A\otimes B}_x = b^A_B b^B_x, \qquad d^{A\otimes B}_x = b^A_B d^B_x
\end{equation}

\begin{equation}
 \label{eq:recommendation2}
u^{A\otimes B}_x = d^A_B + u^A_B + b^A_B u^B_x
\end{equation}
\noindent This operation\footnote{The recommendation operator can become equivalent to Shafer's discounting function \cite{Shafer:1976} as explained in Lioma et al. (2009) \cite{LiomaB:2009}, section  3.2.} is associative but not commutative.

\subsection{Subjective Logic for Polyrepresentation}
\label{ss:SL-preliminaries}
The model of Lioma et al. (2010) uses subjective logic to express the polyrepresentation of user information needs and their context. An analogy is made between opinions (from subjective logic) and query context representations (from polyrepresentation). The opinions express beliefs about the truth of a proposition. In this analogy, the proposition is the original query, and the opinions are the query context representations (see Table \ref{tab:analogies}). The opinion of a query context representation can be seen as the extent to which the query context represents the original query. Then, the combination of multiple query context representations can be modelled as fusing opinions about a proposition.

\begin{table}
\centering
\caption{\label{tab:analogies}Polyrepresentation and subjective logic analogies.}
\scalebox{0.75}{
\begin{tabular}{l|l} 
\hline
POLYREPRESENTATION  &SUBJECTIVE LOGIC\\
\hline
$\cdot$ original query &$\cdot$ proposition\\
\hline
$\cdot$ query context   & $\cdot$ opinion\\
\hline
$\cdot$ how well a query context   & $\cdot$ belief, disbelief \& uncertainty of an  \\
represents the original query&opinion about the truth of the proposition\\
\hline
$\cdot$ combination of query     &\multirow{2}{*}{$\cdot$ opinion fusion}\\
context representations &\\
\hline
\end{tabular}
}
\end{table}

\section{Practical Application}
\label{s:Experiments}
In this paper, we implement and experimentally test the model of Lioma et al. (2010) summarised in the previous section. Given an IR test collection with rich query context representations, we convert these representations into subjective logic opinions and we produce their combinations using the consensus and recommendation operations. The strength of each combination is computed as a probability using Equation \ref{eq:expectation}, so that the higher the probability, the closer a combination represents the original query. To evaluate this model, we conduct retrieval experiments using the same query context combinations (completely independently from the subjective logic computations). If the best performing combination of query context (according to retrieval performance) corresponds to the combination of subjective logic opinions with the highest probability, we consider the model validated. In this work, we focus only on pairwise combinations of query context representations. 

Section \ref{ss:Settings} describes the dataset and settings used in the polyrepresentation and retrieval experiments. Section \ref{ss:Experiments} presents the experimental findings.

\subsection{Experimental Setup}
\label{ss:Settings}

\subsubsection{Dataset and query context representations}
\label{sss:Dataset}
We use the iSearch test collection \cite{LykkeLLI10}, which consists of 46GB of scientific documents from the physics domain. iSearch comes with a set of 65 queries and their relevance assessments, which have been created by 23 lecturers and experienced postgraduate and graduate students from three different university departments of physics. The queries represent real information seeking tasks. The relevance assessment of each query was made by the same user who formulated that query, by examining a pool of documents retrieved for that query. 

Each iSearch query contains the following five representations of different aspects of the user's information need and context: 
\begin{enumerate}
\item user's verbose description of the information sought 
\item background of the user's task
\item description of the user's current work task 
\item description of the user's ideal answer 
\item query terms (keywords) that the user might use in a search engine
\end{enumerate} 

In this work, we consider the keywords (representation no. 5) as the original query, and representations 1-4 as expressions of query context. We choose to regard the keywords as the representation closest to original user query for two reasons. First, the keywords are the most similar to what a user might submit to a search engine, e.g. they are most often in the form of a few key terms or phrases, and are the least verbose of the five representations. Second, earlier results on the iSearch collection indicate \cite{Sorensen} that the single best performing representation among these five is the keywords, indicating that this representation works well with current retrieval models. 

\subsubsection{Retrieval settings}
 We use the Indri IR system\footnote{http://www.lemurproject.org/indri/}. For ranking we use two versions of the language model: with Dirichlet (Dir) and Jelinek-Mercer (JM) smoothing \cite{Croft:2003}. We tune their parameters using the tuning range in Zhai and Lafferty (2002) \cite{ZhaiL:2002}: 
 \begin{itemize}
 \item DIR's $\mu$ $\in$ \{100, 500, 800, 1000, 2000, 3000, 4000, 5000, 8000, 10000\} 
 \item JM's $\lambda$ $\in$ \{0.01, 0.05, 0.1, 0.2, 0.3, 0.4, 0.5, 0.6, 0.7, 0.8, 0.9, 0.95, 0.99\}
 \end{itemize} 
 We report the best retrieval performance separately for Mean Average Precision (MAP), Normalised Discounted Cumulated Gain (NDCG)\footnote{with the following gain values: very relevant = 3, fairly relevant = 2, marginally relevant = 1, non-relevant = 0.}, and Binary Preference (BPREF) in the top 1000 retrieved documents, and also for Precision at 10 (P@10), NDCG at 10 (NDCG@10) and Mean Reciprocal Rank (MRR). These measures contribute different aspects to the evaluation: Both MAP and BPREF measure the average precision of a ranked list, but BPREF differs from MAP because it does not treat non-assessed documents as explicitly non-relevant (whereas MAP does) \cite{BuckleyV04}. NDCG measures the gain of a document based on its position in the result list. The gain is accumulated from the top of the ranked list to the bottom, with the gain of each document discounted at lower ranks. This gain is relative to the ideal based on a known recall base of relevance assessments \cite{JarvelinK02}. P@10 and NDCG@10 focus on the early precision of the top 10 retrieved documents. MRR \cite{VoorheesT99} corresponds to the multiplicative inverse of the rank of the first relevant document retrieved, i.e. it focuses on the retrieval quality of the very top of the ranked list.

\subsection{Experimental Findings}
\label{ss:Experiments}
\def\firstcircle{(90:0.5cm) circle (0.8cm)}
\def\secondcircle{(210:0.5cm) circle (0.8cm)}
\def\thirdcircle{(330:0.5cm) circle (0.8cm)}
\begin{figure*}
\centering
\begin{tikzpicture}
\node[anchor=north,text width=14cm,text justified] () at (0,0)
   {\textbf{The user query ($Q$) and two representations ($A$ and $B$) of its context. Each representation contains positive and negative evidence about the strength of the combination $A \oplus B$ or $A \otimes B$ with respect to the query.}}; 
\end{tikzpicture}
\begin{minipage}[b]{0.45\linewidth}
\centering
\begin{tikzpicture}
\begin{scope}
\clip \thirdcircle;
\fill[cyan] \secondcircle;
\end{scope}
\begin{scope}
\clip \firstcircle;
\fill[cyan] \secondcircle;
\end{scope}
\draw \firstcircle node[text=black,above] {\tiny Query $Q$};
\draw \secondcircle node [text=black,below left] {\tiny $A$};
\draw \thirdcircle node [text=black,below right] {\tiny $B$};
\end{tikzpicture}
\caption{\label{fig:1}Positive evidence of $A$ (in $A \oplus B$).}
\end{minipage}
\hspace{0.5cm}
\begin{minipage}[b]{0.45\linewidth}
\centering
\begin{tikzpicture}
\begin{scope}
\clip \secondcircle;
\fill[cyan] \thirdcircle;
\end{scope}
\begin{scope}
\clip \firstcircle;
\fill[cyan] \thirdcircle;
\end{scope}
\draw \firstcircle node[text=black,above] {\tiny Query $Q$};
\draw \secondcircle node [text=black,below left] {\tiny $A$};
\draw \thirdcircle node [text=black,below right] {\tiny $B$};
\end{tikzpicture}
\caption{\label{fig:2}Positive evidence of $B$ (in $A \oplus B$).}
\end{minipage}
\hspace{0.5cm}
\begin{minipage}[b]{0.45\linewidth}
\centering
\begin{tikzpicture}
\begin{scope}[shift={(1cm,0cm)}]
\begin{scope}[even odd rule]
\clip \thirdcircle (-1.3,-1.3) rectangle (1.3,1.3);
\clip \firstcircle (-1.3,-1.3) rectangle (1.3,1.3);
\fill[cyan] \secondcircle;
\end{scope}
\draw \firstcircle node[text=black,above] {\tiny Query $Q$};
\draw \secondcircle node [text=black,below left] {\tiny $A$};
\draw \thirdcircle node [text=black,below right] {\tiny $B$};
\end{scope}
\end{tikzpicture}
\caption{\label{fig:3}Negative evidence of $A$ (in $A \oplus B$).}
\end{minipage}
\hspace{0.5cm}
\begin{minipage}[b]{0.45\linewidth}
\centering
\begin{tikzpicture}
\begin{scope}[shift={(1cm,0cm)}]
\begin{scope}[even odd rule]
\clip \secondcircle (-1.3,-1.3) rectangle (1.3,1.3);
\clip \firstcircle (-1.3,-1.3) rectangle (1.3,1.3);
\fill[cyan] \thirdcircle;
\end{scope}
\draw \firstcircle node[text=black,above] {\tiny Query $Q$};
\draw \secondcircle node [text=black,below left] {\tiny $A$};
\draw \thirdcircle node [text=black,below right] {\tiny $B$};
\end{scope}
\end{tikzpicture}
\caption{\label{fig:4}Negative evidence of $B$ (in $A \oplus B$).}
\end{minipage}
\hspace{0.5cm}
\begin{minipage}[b]{0.45\linewidth}
\centering
\begin{tikzpicture}
\begin{scope}
\clip \thirdcircle;
\fill[cyan] \secondcircle;
\end{scope}
\draw \firstcircle node[text=black,above] {\tiny Query $Q$};
\draw \secondcircle node [text=black,below left] {\tiny $A$};
\draw \thirdcircle node [text=black,below right] {\tiny $B$};
\end{tikzpicture}
\caption{\label{fig:5}Positive evidence of $A$ about $B$ and also of $B$ about $A$ (in $A \otimes B$).}
\end{minipage}
\end{figure*}

\subsubsection{Mapping Opinions to Evidence}
\label{ss:EvidenceSpace}
According to the model of Lioma et al. (2010) summarised in section \ref{s:Model}, each representation of query context corresponds to an opinion about the original query. The belief, disbelief and uncertainty of these opinions must be computed from features of these query context representations (completely independently of the retrieval process). These features are referred to as evidence, and can be either positive or negative with respect to the original query, depending on whether they support it or not. Subjective logic maps this type of evidence to opinions as follows \cite{Josang:2001}. Let $r$ denote positive evidence, and $s$ denote negative evidence. Then, the correspondence between this evidence and the belief, disbelief, and uncertainty $b,d,u$ is: 
\begin{equation}
\label{eq:mapping}
b = \frac{r}{r+s+2} \qquad d=\frac{s}{r+s+2} \qquad u = \frac{2}{r+s+2}
\end{equation}

What constitutes positive and negative evidence can be defined in various ways. In this work, we use the terms of the query context representations in a simple bag of words approach (regardless of their frequency, co-occurrence, grammatical or any other feature). We consider each query context representation and the original query as sets of terms. Then, positive evidence is the number of terms that occur in overlaps of these sets, and negative evidence is the number of terms that occur in complements of these sets, as described below. 

\begin{table*}
\centering
\caption{\label{tab:polyrep}Pairwise polyrepresentation of query context representations through consensus ($\oplus$) and recommendation ($\otimes$) operations. Consensus and recommendation reflect the strength of each combination as a probability: the higher the probability, the more likely that the combination will benefit retrieval performance. Recommendation is non-commutative, so $A \otimes B$ and $B \otimes A$ denote the combination order.}
\begin{tabular}{l|c|c|c} 
\multicolumn{4}{c}{$ $}\\
\hline
\multicolumn{4}{c}{POLYREPRESENTATION PROBABILITIES}\\ 
\hline
\multicolumn{4}{c}{(I) No Pre-Processing}\\ 
\hline
 \multirow{2}{*}{Query Context Representations}  &Consensus  &\multicolumn{2}{c}{Recommendation}\\ 
    &$A \oplus B$  &$A \otimes B$    &$B \otimes A$\\ 
\hline
user background - ideal answer      &0.1850             &0.4126             &0.4192\\
user background - work task       &0.1965             &0.3527             &0.4518\\
information need - user background  &0.1862             &0.3759             &0.4594\\
information need - ideal answer     &\bf0.2359          &0.3687             &0.4450\\
information need - work task      &0.2064             &0.3096             &\bf0.4764\\
work task - ideal answer          &0.1769             &\bf0.4258          &0.3859\\
\hline
\multicolumn{4}{c}{$ $}\\
\end{tabular}
\begin{tabular}{l|c|c|c} 
\hline
\multicolumn{4}{c}{(II) Lower-Case, No Punctuation}\\ 
\hline
 \multirow{2}{*}{Query Context Representations}  &Consensus  &\multicolumn{2}{c}{Recommendation}\\ 
    &$A \oplus B$  &$A \otimes B$    &$B \otimes A$\\ 
\hline
user background - ideal answer      &0.2285             &0.3967             &0.4153\\
user background - work task       &0.2347             &0.3319             &0.4540\\
information need - user background  &0.2426             &0.3470             &0.4702\\
information need - ideal answer     &\bf0.2939          &0.3379             &0.4584\\
information need - work task      &0.2534             &0.2680             &\bf0.4895\\
work task - ideal answer          &0.2123             &\bf0.4137          &0.3811\\
\hline
\multicolumn{4}{c}{$ $}\\
\end{tabular}
\begin{tabular}{l|c|c|c} 
\hline
\multicolumn{4}{c}{(III) Lower-Case, No Punctuation, No Stopwords}\\ 
\hline
 \multirow{2}{*}{Query Context Representations}  &Consensus  &\multicolumn{2}{c}{Recommendation}\\ 
    &$A \oplus B$  &$A \otimes B$    &$B \otimes A$\\ 
\hline
user background - ideal answer      &0.1972             &0.4365             &0.4499\\
user background - work task       &0.1826             &0.4030             &0.4641\\
information need - user background  &0.2457             &0.3928             &0.4787\\
information need - ideal answer     &\bf0.3154          &0.3493             &0.4734\\
information need - work task      &0.2715             &0.2991             &\bf0.4907\\
work task - ideal answer          &0.1902             &\bf0.4436          &0.4274\\
\hline
\multicolumn{4}{c}{$ $}\\
\end{tabular}
\begin{tabular}{l|c|c|c} 
\hline
\multicolumn{4}{c}{(IV) Lower-Case, No Punctuation, No Stopwords, Stemming}\\ 
\hline
 \multirow{2}{*}{Query Context Representations}  &Consensus  &\multicolumn{2}{c}{Recommendation}\\ 
    &$A \oplus B$  &$A \otimes B$    &$B \otimes A$\\ 
\hline
user background - ideal answer        &0.2278             &0.4211             &0.4424\\
user background - work task         &0.2203             &0.3771             &0.4616\\
information need - user background    &0.2885             &0.3607             &0.4837\\
information need - ideal answer       &\bf0.3674          &0.3138             &0.4868\\
information need - work task        &0.3106             &0.2615             &\bf0.5027\\
work task - ideal answer            &0.2172             &\bf0.4307          &0.4170\\
\hline
\multicolumn{4}{c}{$ $}\\
\end{tabular}
\end{table*}

For the consensus of two query context representations $A$ and $B$ ($A \oplus B$) with respect to a user query $Q$, $A$'s positive evidence is the number of terms occurring in $A \cap B$ and also in $A \cap Q$ (Figure \ref{fig:1}). Similarly, $B$'s positive evidence is the number of terms occurring in $A \cap B$ and also in $B \cap Q$ (Figure \ref{fig:2}). $A$'s negative evidence is the number of terms occurring in the complement of $A - B$ and $A - Q$ (Figure \ref{fig:3}). $B$'s negative evidence is the number of terms occurring in the complement of $B - A$ and $B - Q$ (Figure \ref{fig:4}).

For the recommendation of $A$ to $B$ and of $B$ to $A$ ($A \otimes B$), both $A$'s and $B$'s positive evidence is the number of terms occurring in $A \cap B$ (Figure \ref{fig:5}). The negative evidence of $A$ and $B$ is the same as in the consensus above. 

We apply these separate pre-processing options when counting terms: 
\begin{itemize}
\item[](I) no-preprocessing at all
\item[](II) lower-case, punctuation removal
\item[](III) II + stopword removal
\item[](IV) III + stemming
\end{itemize}
We use the SMART stopword list, as appears in the appendices of Lewis et al. (2004) \cite{LewisYRL04}, and the Porter stemming algorithm \cite{Porter80}.

Any type and amount of contextual information that can potentially be useful may be used as evidence, for instance statistical, linguistic, algorithmic or other features of the queries (e.g. see \cite{LiomaB:2009} for pragmatic query features). Additional evidence can be modelled by introducing more opinions. 

Having collected this evidence from the term statistics of the query context representations, we compute their combinations by consensus (using Equations \ref{eq:consensus1} - \ref{eq:consensus2}) and recommendation (using Equations \ref{eq:recommendation1} - \ref{eq:recommendation2}). The output of these equations is fused beliefs, disbeliefs, and uncertainties for each combination, which we feed into Equation \ref{eq:expectation} to compute the final strength of each combination - we refer to this as \textit{polyrepresentation probability}. In Equation \ref{eq:expectation}, we set the prior $\alpha = 0.5$ assuming a binary frame of discernment (i.e. having two states and dividing $\alpha$ uniformally across these).

\subsubsection{Polyrepresentation Findings}
\label{ss:Results1}

Table \ref{tab:polyrep} shows the probabilities of each combination of query context representations computed as described above. The three query context representations that give higher polyrepresentation probabilities consistently for all preprocessing options are the user's information need, work task and ideal answer. Among these, work task and information need give the highest polyrepresentation probability at all times, using the recommendation operation (in this order). The assumption behind this combination is that (a) work task and information need are dependent, and that (b) the work task is a better contextual representation of the information need than of the original query. This can be seen in the very low probability fetched by the recommendation information need $\otimes$ work task (which is the lowest probability for all recommendation combinations). Simply speaking, what we see here is two query context representations that are potentially valuable to retrieval effectiveness but only if their effect is channelled in a specific way so that their effect is traced according to their dependence. Their dependence in this case is that the work task context supports the information need, and the information need context supports the query. This is in line with the cognitive interpretation of user's work task in interactive IR laid down in Ingwersen (1996) \cite{Ingwersen:1996}.
  
Table \ref{tab:polyrep} also shows that recommendation results in higher overall polyrepresentation probabilities than consensus, i.e. combinations that are likely to be more reliable. To assess the predictions of the query context combinations in Table \ref{tab:polyrep}, we compare them to the actual retrieval performance of each combination, described next. No part of the retrieval process has been informed by the polyrepresentation computation described above, and vice versa.

\begin{table*}
\centering
\caption{\label{tab:scores}Retrieval performance using the language model with Dirichlet (DIR) and Jelinek-Mercer (JM) smoothing. The best combination of query context representations is shown in bold. $\dagger$ marks the best overall score per evaluation measure.}
\scalebox{1}{
\begin{tabular}{l|cc|cc|cc} 
\hline
\multirow{2}{*}{Query Context Representations}    &\multicolumn{2}{c|}{MAP}    &\multicolumn{2}{|c|}{NDCG}   &\multicolumn{2}{|c}{BPREF}\\ 
&DIR&JM&DIR&JM&DIR&JM\\
\hline
user background - ideal answer      &0.0588&0.0703  &0.1928&0.2217  &0.2042&0.2347  \\
user background - work task       &0.0933&0.1095  &0.2589&0.3062  &0.2737&0.3059  \\
information need - user background  &0.1073&0.1229  &0.2919&0.3341  &0.2844&0.3213  \\
information need - ideal answer     &0.0945&0.1077  &0.2670&0.2983  &0.2508&0.2834  \\
information need - work task      &\bf0.1175&\bf0.1310$\dagger$  &\bf0.3117&\bf0.3563  &\bf0.2974&\bf0.3270  \\
work task - ideal answer          &0.0849&0.0925  &0.2489&0.2895  &0.2480&0.2960  \\
\hline
original query (no context)           &0.1156&0.1268  &0.3339&0.3572$\dagger$  &0.3201&0.3340$\dagger$\\
\hline
\multicolumn{4}{c}{$ $}\\
\end{tabular}
}
\scalebox{1}{
\begin{tabular}{l|cc|cc|cc} 
\hline
\multirow{2}{*}{Query Context Representations}    &\multicolumn{2}{c|}{P@10} &\multicolumn{2}{|c|}{NDCG@10}    &\multicolumn{2}{|c}{MRR}\\ 
&DIR&JM&DIR&JM&DIR&JM\\
\hline
user background - ideal answer      &0.1446&0.1477  &0.1298&0.1314  &0.3069&0.3310\\
user background - work task       &0.2077&0.2277  &0.1911&0.2061  &0.3884&0.3907\\
information need - user background  &0.2154&0.2138  &0.2004&0.2075  &0.4222&0.4600\\
information need - ideal answer     &0.2185&0.2154  &0.1976&0.1919  &0.4394&0.4167\\
information need - work task      &\bf0.2523&\bf0.2554$\dagger$  &\bf0.2246&\bf0.2352$\dagger$  &\bf0.4582&\bf0.4880\\
work task - ideal answer          &0.2000&0.1892  &0.1883&0.1805  &0.4134&0.3997\\
\hline
original query (no context)           &0.2492&0.2431  &0.2287&0.2175  &0.5267$\dagger$&0.4958\\
\hline
\end{tabular}
}
\end{table*}

\begin{figure*}
\centering
\begin{tikzpicture}[baseline,scale=0.55]
			\begin{axis}[
			font=\large,			
			title=,
			xlabel=Polyrepresentation Probability,
			ylabel=MAP
			]
			\pgfplotstableread{polyrep-ret.data}\table 
			\addplot[mark=+] table[x index=0,y index=1] from \table;
\end{axis}
\end{tikzpicture}
\centering
\begin{tikzpicture}[baseline,scale=0.55]
			\begin{axis}[
			font=\large,			
			title=,
			xlabel=Polyrepresentation Probability,
			ylabel=NDCG
			]
			\pgfplotstableread{polyrep-ret.data}\table 
			\addplot[mark=+] table[x index=0,y index=2] from \table;
\end{axis}
\end{tikzpicture}
\centering
\begin{tikzpicture}[baseline,scale=0.55]
			\begin{axis}[
			font=\large,			
			title=,
			xlabel=Polyrepresentation Probability,
			ylabel=BPREF
			]
			\pgfplotstableread{polyrep-ret.data}\table 
			\addplot[mark=+] table[x index=0,y index=3] from \table;
\end{axis}
\end{tikzpicture}
\begin{tikzpicture}
\node[anchor=north,text width=16.5cm] () at (0,0)
   {\textbf{$\qquad$}}; 
\end{tikzpicture}
\begin{tikzpicture}
\node[anchor=north,text width=16.5cm] () at (0,0)
   {\textbf{$\qquad$}}; 
\end{tikzpicture}
\begin{tikzpicture}
\node[anchor=north,text width=16.5cm] () at (0,0)
   {\textbf{$\qquad$}}; 
\end{tikzpicture}
\centering
\begin{tikzpicture}[baseline,scale=0.55]
			\begin{axis}[
			font=\large,			
			title=,
			xlabel=Polyrepresentation Probability,
			ylabel=P@10
			]
			\pgfplotstableread{polyrep-ret.data}\table 
			\addplot[mark=+] table[x index=0,y index=4] from \table;
\end{axis}
\end{tikzpicture}
\centering
\begin{tikzpicture}[baseline,scale=0.55]
			\begin{axis}[
			font=\large,			
			title=,
			xlabel=Polyrepresentation Probability,
			ylabel=NDCG@10
			]
			\pgfplotstableread{polyrep-ret.data}\table 
			\addplot[mark=+] table[x index=0,y index=5] from \table;
\end{axis}
\end{tikzpicture}
\centering
\begin{tikzpicture}[baseline,scale=0.55]
			\begin{axis}[
			font=\large,			
			title=,
			xlabel=Polyrepresentation Probability,
			ylabel=MRR
			]
			\pgfplotstableread{polyrep-ret.data}\table 
			\addplot[mark=+] table[x index=0,y index=6] from \table;
\end{axis}
\end{tikzpicture}
\caption{\label{fig:comparison}Probability of each polyrepresentation combination (x axis) vs. retrieval performance with JM (y axis) for all query context representation combinations (the points correspond to the scores in Tables \ref{tab:polyrep} \& \ref{tab:scores}).}
\end{figure*}
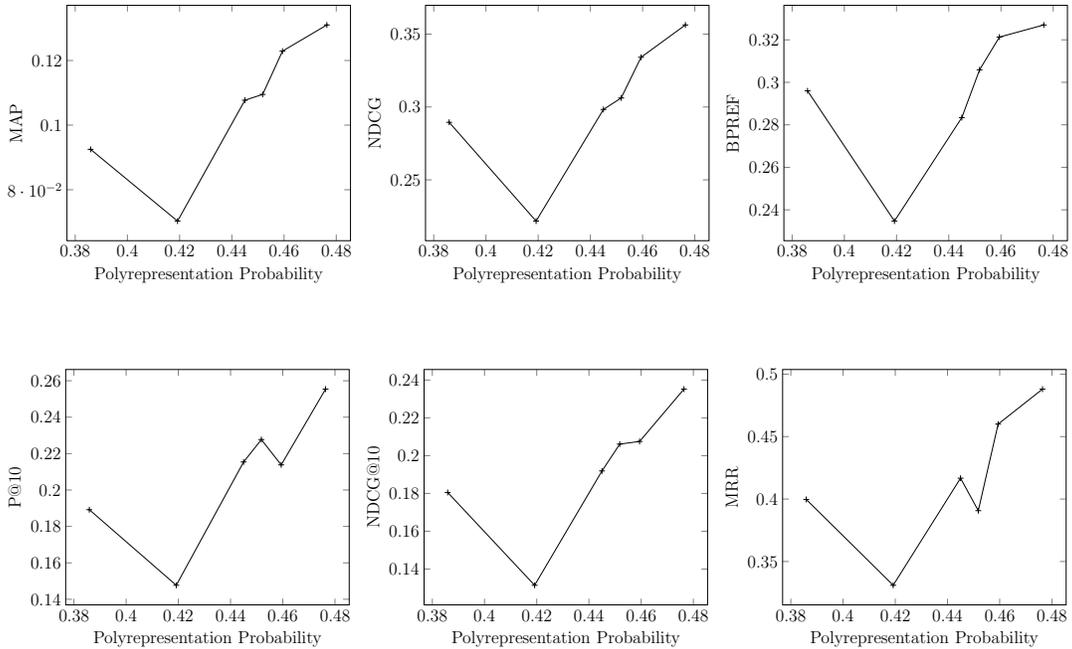

\subsubsection{Retrieval Findings}
Table \ref{tab:scores} shows the retrieval performance of the combinations of query context representations we saw in Table \ref{tab:polyrep}. For these runs, we use the text of each query context representation as query text, and the same four types of preprocessing reported in Section \ref{ss:EvidenceSpace}. We do not  weight separately any of the query fields; we simply concatenate all text into one query. Table \ref{tab:scores} also displays the retrieval performance when using the original query without any context, for reference (there is no polyrepresentation of context in this run). We see that the combination with the highest polyrepresentation probability in Table \ref{tab:polyrep}, information need and work task, gives also the best retrieval performance among all combinations. This is consistent for all six evaluation measures, and for both retrieval models. Comparing these results to the performance of the original query without context, we see that the best polyrepresentation run improves retrieval performance over the original query for MAP, P@10 and NDCG@10, and is comparable to the original query for the remaining evaluation measures. Overall, these findings indicate that the polyrepresentation probabilities tend to agree with observed retrieval performance. This agreement is also visually displayed in Figure \ref{fig:comparison}. We see that there is overall consistency among the two, apart from the case of background - work task. 

The scores shown in Table \ref{tab:scores} are averaged over all queries, meaning that they can be affected by outliers. Figures \ref{fig:perquery1} - \ref{fig:perquery2} present a detailed per-query overview of the retrieval performance of each query (measured in MAP only for DIR) against the belief and uncertainty of the query context representations combined with consensus and recommendation\footnote{For the recommendation, we present one order of combination (out of the two); we can report that the other combination order follows very similar trends.}. This belief and uncertainty are the ingredients used to compute the polyrepresentation probabilities with Equation \ref{eq:expectation}. Hence, these figures serve to explain how the polyrepresentatin probabilities can be decomposed and how their components correlate to retrieval performance. The correlation between MAP and the respective belief and uncertainty values is reported using Spearman's rank correlation coefficient. We can report similar trends for runs with JM and for the other evaluation measures that average retrieval precision in the top 1000 documents.

Figures \ref{fig:perquery1} - \ref{fig:perquery2} show that the combination of information need and work task (which gave the highest polyrepresentation probability and also the highest retrieval performance) does not have the highest correlation with MAP on a per query basis, neither for belief nor for uncertainty. The highest correlation with respect to these is given by the combination of information need with user background. However, the combination of information need and work task has overall the highest belief values among all combinations. This shows that this combination has somewhat stronger evidence than the others, which may not always correlate strongly with the MAP score flunctuations for each query, but which nevertheless is an overall better representation of the query context.

\begin{figure*}
\centering
\begin{tikzpicture}
\node[anchor=north,text width=14cm,text justified] () at (0,0)
   {\textbf{The belief and uncertainty (y axis) of the representations used for the consensus combination (this page) and the recommendation operator (next page) against MAP (x axis) per query, and their rank correlation coefficients (Spearman's $\rho$). The order of the combination for recommendation is shown in the title of each figure.}}; 
\end{tikzpicture}
\centering
\pgfplotsset{every axis label/.append style={font=\large}}
\tikzset{every mark/.append style={font=\large}}
\begin{minipage}[b]{0.3\linewidth}
\centering
\begin{tikzpicture}[baseline,scale=0.55]
			\begin{axis}[
			font=\large,			
			title=Background $\oplus$ Ideal Answer,
			xlabel=MAP,
			ylabel=Belief,
			yticklabels={0.002,0.1},
			ytick={0.002,0.1}			
			]
			\pgfplotstableread{BK-IA-DIR-MAP}\table 
			\addplot[only marks,mark=+] table[x index=2,y index=4] from \table;
\end{axis}
\end{tikzpicture}
\caption{\label{fig:perquery1}$\rho=0.123$}
\end{minipage}
\begin{minipage}[b]{0.3\linewidth}
\centering
\begin{tikzpicture}[baseline,scale=0.55]
			\begin{axis}[
			font=\large,			
			title=Background $\oplus$ Task,
			xlabel=MAP,
			ylabel=Belief,
			yticklabels={0.01,0.1},
			ytick={0.005,0.1}			
			]
			\pgfplotstableread{BK-WT-DIR-MAP}\table 
			\addplot[only marks,mark=+] table[x index=2,y index=4] from \table;
\end{axis}
\end{tikzpicture}
\caption{$\rho=-0.014$}
\end{minipage}
\begin{minipage}[b]{0.3\linewidth}
\centering
\begin{tikzpicture}[baseline,scale=0.55]
			\begin{axis}[
			font=\large,			
			title=Information need $\oplus$ Background,
			xlabel=MAP,
			ylabel=Belief,
			yticklabels={0.09,0.1},
			ytick={0.009,0.1}			
			]
			\pgfplotstableread{IN-BK-DIR-MAP}\table 
			\addplot[only marks,mark=+] table[x index=2,y index=4] from \table;
\end{axis}
\end{tikzpicture}
\caption{$\rho=0.483$}
\end{minipage}
\begin{tikzpicture}
\node[anchor=north,text width=16.5cm] () at (0,0)
   {\textbf{$\qquad$}}; 
\end{tikzpicture}
\begin{tikzpicture}
\node[anchor=north,text width=16.5cm] () at (0,0)
   {\textbf{$\qquad$}}; 
\end{tikzpicture}
\begin{tikzpicture}
\node[anchor=north,text width=16.5cm] () at (0,0)
   {\textbf{$\qquad$}}; 
\end{tikzpicture}
\begin{minipage}[b]{0.3\linewidth}
\centering
\begin{tikzpicture}[baseline,scale=0.55]
			\begin{axis}[
			font=\large,			
			title=Information Need $\oplus$ Ideal Answer,
			xlabel=MAP,
			ylabel=Belief,
			yticklabels={0.02,0.1},
			ytick={0.002,0.1},			
			xtick={0.0005,0.02},
			xticklabels={0.0005,0.02}
			]
			\pgfplotstableread{IN-IA-DIR-MAP}\table 
			\addplot[only marks,mark=+] table[x index=2,y index=4] from \table;
\end{axis}
\end{tikzpicture}
\caption{$\rho=0.411$}
\end{minipage}
\begin{minipage}[b]{0.3\linewidth}
\centering
\begin{tikzpicture}[baseline,scale=0.55]
			\begin{axis}[
			font=\large,			
			title=Information Need $\oplus$ Task,
			xlabel=MAP,
			ylabel=Belief,
			yticklabels={0.2,0.05},
			ytick={0.05,0.2}			
			]
			\pgfplotstableread{IN-WT-DIR-MAP}\table 
			\addplot[only marks,mark=+] table[x index=2,y index=4] from \table;
\end{axis}
\end{tikzpicture}
\caption{$\rho=0.359$}
\end{minipage}
\begin{minipage}[b]{0.3\linewidth}
\centering
\begin{tikzpicture}[baseline,scale=0.55]
			\begin{axis}[
			font=\large,			
			title=Task $\oplus$ Ideal Answer,
			xlabel=MAP,
			ylabel=Belief,
			yticklabels={0.02,0.005},
			ytick={0.02,0.005}			
			]
			\pgfplotstableread{WT-IA-DIR-MAP}\table 
			\addplot[only marks,mark=+] table[x index=2,y index=4] from \table;
\end{axis}
\end{tikzpicture}
\caption{$\rho=0.049$}
\end{minipage}
\begin{tikzpicture}
\node[anchor=north,text width=16.5cm] () at (0,0)
   {\textbf{$\qquad$}}; 
\end{tikzpicture}
\begin{tikzpicture}
\node[anchor=north,text width=16.5cm] () at (0,0)
   {\textbf{$\qquad$}}; 
\end{tikzpicture}
\begin{tikzpicture}
\node[anchor=north,text width=16.5cm] () at (0,0)
   {\textbf{$\qquad$}}; 
\end{tikzpicture}
\begin{minipage}[b]{0.3\linewidth}
\centering
\begin{tikzpicture}[baseline,scale=0.55]
			\begin{axis}[
			font=\large,			
			title=Background $\oplus$ Ideal Answer,
			xlabel=MAP,
			ylabel=Uncertainty
			]
			\pgfplotstableread{BK-IA-DIR-MAP}\table 
			\addplot[only marks,mark=+] table[x index=2,y index=5] from \table;
\end{axis}
\end{tikzpicture}
\caption{$\rho=-0.009$}
\end{minipage}
\begin{minipage}[b]{0.3\linewidth}
\centering
\begin{tikzpicture}[baseline,scale=0.55]
			\begin{axis}[
			font=\large,			
			title=Background $\oplus$ Task,
			xlabel=MAP,
			ylabel=Uncertainty
			]
			\pgfplotstableread{BK-WT-DIR-MAP}\table 
			\addplot[only marks,mark=+] table[x index=2,y index=5] from \table;
\end{axis}
\end{tikzpicture}
\caption{$\rho=-0.198$}
\end{minipage}
\begin{minipage}[b]{0.3\linewidth}
\centering
\begin{tikzpicture}[baseline,scale=0.55]
			\begin{axis}[
			font=\large,			
			title=Information need $\oplus$ Background,
			xlabel=MAP,
			ylabel=Uncertainty
			]
			\pgfplotstableread{IN-BK-DIR-MAP}\table 
			\addplot[only marks,mark=+] table[x index=2,y index=5] from \table;
\end{axis}
\end{tikzpicture}
\caption{$\rho=-0.209$}

\end{minipage}
\begin{tikzpicture}
\node[anchor=north,text width=16.5cm] () at (0,0)
   {\textbf{$\qquad$}}; 
\end{tikzpicture}
\begin{tikzpicture}
\node[anchor=north,text width=16.5cm] () at (0,0)
   {\textbf{$\qquad$}}; 
\end{tikzpicture}
\begin{tikzpicture}
\node[anchor=north,text width=16.5cm] () at (0,0)
   {\textbf{$\qquad$}}; 
\end{tikzpicture}

\begin{minipage}[b]{0.3\linewidth}
\centering
\begin{tikzpicture}[baseline,scale=0.55]
			\begin{axis}[
			font=\large,			
			title=Information Need $\oplus$ Ideal Answer,
			xlabel=MAP,
			ylabel=Uncertainty,
			xtick={0.0005,0.02},
			xticklabels={0.0005,0.02}
			]
			\pgfplotstableread{IN-IA-DIR-MAP}\table 
			\addplot[only marks,mark=+] table[x index=2,y index=5] from \table;
\end{axis}
\end{tikzpicture}
\caption{$\rho=-0.146$}
\end{minipage}
\begin{minipage}[b]{0.3\linewidth}
\centering
\begin{tikzpicture}[baseline,scale=0.55]
			\begin{axis}[
			font=\large,			
			title=Information Need $\oplus$ Task,
			xlabel=MAP,
			ylabel=Uncertainty
			]
			\pgfplotstableread{IN-WT-DIR-MAP}\table 
			\addplot[only marks,mark=+] table[x index=2,y index=5] from \table;
\end{axis}
\end{tikzpicture}
\caption{$\rho=-0.065$}
\end{minipage}
\begin{minipage}[b]{0.3\linewidth}
\centering
\begin{tikzpicture}[baseline,scale=0.55]
			\begin{axis}[
			font=\large,			
			title=Task $\oplus$ Ideal Answer,
			xlabel=MAP,
			ylabel=Uncertainty
			]
			\pgfplotstableread{WT-IA-DIR-MAP}\table 
			\addplot[only marks,mark=+] table[x index=2,y index=5] from \table;
\end{axis}
\end{tikzpicture}
\caption{$\rho=-0-007$}
\end{minipage}
\label{fig:perquery}
\end{figure*}
\begin{figure*}
\centering
\pgfplotsset{every axis label/.append style={font=\large}}
\tikzset{every mark/.append style={font=\large}}
\begin{minipage}[b]{0.3\linewidth}
\centering
\begin{tikzpicture}[baseline,scale=0.55]
			\begin{axis}[
			font=\large,			
			title=Ideal Answer $\otimes$ Background,
			xlabel=MAP,
			ylabel=Belief,
			yticklabels={0.002,0.1},
			ytick={0.002,0.1}			
			]
			\pgfplotstableread{BK-IA-DIR-MAP}\table 
			\addplot[only marks,mark=+] table[x index=2,y index=6] from \table;
\end{axis}
\end{tikzpicture}
\caption{$\rho=0.062$}
\end{minipage}
\begin{minipage}[b]{0.3\linewidth}
\centering
\begin{tikzpicture}[baseline,scale=0.55]
			\begin{axis}[
			font=\large,			
			title=Task $\otimes$ Background,
			xlabel=MAP,
			ylabel=Belief,
			yticklabels={0.01,0.1},
			ytick={0.005,0.1}			
			]
			\pgfplotstableread{BK-WT-DIR-MAP}\table 
			\addplot[only marks,mark=+] table[x index=2,y index=6] from \table;
\end{axis}
\end{tikzpicture}
\caption{$\rho=0.044$}
\end{minipage}
\begin{minipage}[b]{0.3\linewidth}
\centering
\begin{tikzpicture}[baseline,scale=0.55]
			\begin{axis}[
			font=\large,			
			title=Background $\otimes$ Information Need,
			xlabel=MAP,
			ylabel=Belief,
			yticklabels={0.09,0.1},
			ytick={0.009,0.1}			
			]
			\pgfplotstableread{IN-BK-DIR-MAP}\table 
			\addplot[only marks,mark=+] table[x index=2,y index=6] from \table;
\end{axis}IN-BK-DIR-MAP
\end{tikzpicture}
\caption{$\rho=0.238$}
\end{minipage}
\begin{tikzpicture}
\node[anchor=north,text width=16.5cm] () at (0,0)
   {\textbf{$\qquad$}}; 
\end{tikzpicture}
\begin{tikzpicture}
\node[anchor=north,text width=16.5cm] () at (0,0)
   {\textbf{$\qquad$}}; 
\end{tikzpicture}
\begin{tikzpicture}
\node[anchor=north,text width=16.5cm] () at (0,0)
   {\textbf{$\qquad$}}; 
\end{tikzpicture}
\begin{minipage}[b]{0.3\linewidth}
\centering
\begin{tikzpicture}[baseline,scale=0.55]
			\begin{axis}[
			font=\large,			
			title=Ideal $\otimes$ Information Need,
			xlabel=MAP,
			ylabel=Belief,
			yticklabels={0.02,0.1},
			ytick={0.002,0.1}			
			]
			\pgfplotstableread{IN-IA-DIR-MAP}\table 
			\addplot[only marks,mark=+] table[x index=2,y index=6] from \table;
\end{axis}
\end{tikzpicture}
\caption{$\rho=0.086$}
\end{minipage}
\begin{minipage}[b]{0.3\linewidth}
\centering
\begin{tikzpicture}[baseline,scale=0.55]
			\begin{axis}[
			font=\large,			
			title=Task $\otimes$ Information Need,
			xlabel=MAP,
			ylabel=Belief,
			yticklabels={0.05,0.15},
			ytick={0.05,0.15}			
			]
			\pgfplotstableread{IN-WT-DIR-MAP}\table 
			\addplot[only marks,mark=+] table[x index=2,y index=6] from \table;
\end{axis}
\end{tikzpicture}
\caption{$\rho=0.117$}
\end{minipage}
\begin{minipage}[b]{0.3\linewidth}
\centering
\begin{tikzpicture}[baseline,scale=0.55]
			\begin{axis}[
			font=\large,			
			title=Ideal Answer $\otimes$ Task,
			xlabel=MAP,
			ylabel=Belief,
			yticklabels={0.02,0.1},
			ytick={0.02,0.1}			
			]
			\pgfplotstableread{WT-IA-DIR-MAP}\table 
			\addplot[only marks,mark=+] table[x index=2,y index=6] from \table;
\end{axis}
\end{tikzpicture}
\caption{$\rho=0.105$}
\end{minipage}
\begin{tikzpicture}
\node[anchor=north,text width=16.5cm] () at (0,0)
   {\textbf{$\qquad$}}; 
\end{tikzpicture}
\begin{tikzpicture}
\node[anchor=north,text width=16.5cm] () at (0,0)
   {\textbf{$\qquad$}}; 
\end{tikzpicture}
\begin{tikzpicture}
\node[anchor=north,text width=16.5cm] () at (0,0)
   {\textbf{$\qquad$}}; 
\end{tikzpicture}
\begin{minipage}[b]{0.3\linewidth}
\centering
\begin{tikzpicture}[baseline,scale=0.55]
			\begin{axis}[
			font=\large,			
			title=Ideal Answer $\otimes$ Background,
			xlabel=MAP,
			ylabel=Uncertainty
			]
			\pgfplotstableread{BK-IA-DIR-MAP}\table 
			\addplot[only marks,mark=+] table[x index=2,y index=7] from \table;
\end{axis}
\end{tikzpicture}
\caption{$\rho=-0.080$}
\end{minipage}
\begin{minipage}[b]{0.3\linewidth}
\centering
\begin{tikzpicture}[baseline,scale=0.55]
			\begin{axis}[
			font=\large,			
			title=Task $\otimes$ Background,
			xlabel=MAP,
			ylabel=Uncertainty
			]
			\pgfplotstableread{BK-WT-DIR-MAP}\table 
			\addplot[only marks,mark=+] table[x index=2,y index=7] from \table;
\end{axis}
\end{tikzpicture}
\caption{$\rho=0.019$}
\end{minipage}
\begin{minipage}[b]{0.3\linewidth}
\centering
\begin{tikzpicture}[baseline,scale=0.55]
			\begin{axis}[
			font=\large,			
			title=Background $\otimes$ Information Need,
			xlabel=MAP,
			ylabel=Uncertainty
			]
			\pgfplotstableread{IN-BK-DIR-MAP}\table 
			\addplot[only marks,mark=+] table[x index=2,y index=7] from \table;
\end{axis}
\end{tikzpicture}
\caption{$\rho=-0.136$}

\end{minipage}
\begin{tikzpicture}
\node[anchor=north,text width=16.5cm] () at (0,0)
   {\textbf{$\qquad$}}; 
\end{tikzpicture}
\begin{tikzpicture}
\node[anchor=north,text width=16.5cm] () at (0,0)
   {\textbf{$\qquad$}}; 
\end{tikzpicture}
\begin{tikzpicture}
\node[anchor=north,text width=16.5cm] () at (0,0)
   {\textbf{$\qquad$}}; 
\end{tikzpicture}

\begin{minipage}[b]{0.3\linewidth}
\centering
\begin{tikzpicture}[baseline,scale=0.55]
			\begin{axis}[
			font=\large,			
			title=Ideal $\otimes$ Information Need,
			xlabel=MAP,
			ylabel=Uncertainty
			]
			\pgfplotstableread{IN-IA-DIR-MAP}\table 
			\addplot[only marks,mark=+] table[x index=2,y index=7] from \table;
\end{axis}
\end{tikzpicture}
\caption{$\rho=-0.064$}
\end{minipage}
\begin{minipage}[b]{0.3\linewidth}
\centering
\begin{tikzpicture}[baseline,scale=0.55]
			\begin{axis}[
			font=\large,			
			title=Task $\otimes$ Information Need,
			xlabel=MAP,
			ylabel=Uncertainty
			]
			\pgfplotstableread{IN-WT-DIR-MAP}\table 
			\addplot[only marks,mark=+] table[x index=2,y index=7] from \table;
\end{axis}
\end{tikzpicture}
\caption{$\rho=-0.061$}
\end{minipage}
\begin{minipage}[b]{0.3\linewidth}
\centering
\begin{tikzpicture}[baseline,scale=0.55]
			\begin{axis}[
			font=\large,			
			title=Ideal Answer $\otimes$ Task,
			xlabel=MAP,
			ylabel=Uncertainty
			]
			\pgfplotstableread{WT-IA-DIR-MAP}\table 
			\addplot[only marks,mark=+] table[x index=2,y index=7] from \table;
\end{axis}
\end{tikzpicture}
\caption{\label{fig:perquery2}$\rho=-0.136$}
\end{minipage}
\label{fig:perquery}
\end{figure*}

\section{Discussion}
\label{s:Discussion}
The experimental findings presented above can be summarised in three points. 

The experiments indicate that the polyrepresentation predictions regarding the optimal combination of query context representations were in agreement with observed retrieval performance. This validates the model of Lioma et al. (2010) for this dataset and retrieval scenario. Further experiments with other settings and using more than pairwise combinations of representations are needed to ground the model on firmer grounds - we consider this study as a first step in that direction. 

We showed how the components of each polyrepresentation combination were induced from naive bag of words term statistics and converted to belief, disbelief and uncertainty. This was a straight-forward application of the mapping proposed in J\o sang (2001) \cite{Josang:2001} and of features common to IR experimentation. Further features could be used, potentially challenging their mapping to binary (positive or  negative) evidence, resulting in new definitions and mappings of graded features as contextual evidence for IR. 

Finally, we showed how the effect of one representation upon the other can be traced and modelled in a combination. This is an attractive feature that could be used to model and practically adjust the user retrieval experience, through for instance interface functionalities, such as timely interventions or suggestions. 

\section{Conclusions}
\label{s:Conclusions}

This work provided a practical application and analysis of the principle of polyrepresentation formalised using subjective logic, as initially proposed in Lioma et al. (2010). We showed how to map the abstract notions of belief and uncertainty used in that model to real-life evidence drawn from a retrieval test collection, and how to estimate two different types of combinations for polyrepresentation assuming either (a) independence or (b) dependence between the information objects that are combined. Experimental evidence on the polyrepresentation of different types of context relating to user information needs (i.e. work task, user background knowledge, ideal answer) using two state of the art retrieval models, six standand evaluation measures and 65 queries showed that the model of Lioma et al. (2010) can predict their optimal combination prior and independently to the retrieval process.  


\section{Acknowledgments}
We thank the anonymous reviewers for useful feedback.
%
\bibliographystyle{abbrv}
\bibliography{iiix2012}  
%
%

\end{document}